# SACS: A Code Smell Dataset using Semi-automatic Generation Approach

HANYU ZHANG
Department of Industrial and Management Systems Engineering
Waseda University
Japan, Tokyo
bankzhy@akane.waseda.jp

TOMOJI KISHI
Department of Industrial and Management Systems Engineering
Waseda University
Japan, Tokyo
kishi@waseda.jp

***Abstract*— Code smell is a great challenge in software refactoring, which indicates latent design or implementation flaws that may degrade the software maintainability and evolution. Over the past few decades, code smell research has received extensive attention. In particular, the application of machine learning techniques to code smell detection and refactoring has become an increasingly popular research topic in recent years. However, one of the biggest challenges to apply machine learning-technique is the lack of high-quality code smell datasets. Manually constructing such datasets is extremely labor-intensive, as identifying code smells requires substantial development expertise and considerable time investment. In contrast, automatically generated datasets, while scalable, frequently exhibit reduced label reliability and compromised data quality. To overcome this challenge, in this study, we explore a semi-automatic approach to generate a code smell dataset with high quality data samples. Specifically, we first applied a set of automatic generation rules to produce candidate smelly samples. We then employed multiple metrics to group the data samples into an automatically accepted group and a manually reviewed group, enabling reviewers to concentrate their efforts on ambiguous samples. Furthermore, we established structured review guidelines and developed an annotation tool to support the manual validation process. Based on the proposed semi-automatic generation approach, we created an open-source code smell dataset, SACS, covering three widely studied code smells: Long Method, Large Class, and Feature Envy. Each code smell category includes over 10,000 labeled samples. This dataset could provide a large-scale and publicly available benchmark to facilitate future studies on code smell detection and automated refactoring.**

## I. INTRODUCTION

Software Refactoring has been playing an important role in software lifecycle helps to improve the maintainability and readability of the software. It refers to optimizing the software internal structure without changing its external behavior. During the software refactoring process, Code Smell has always been an important topic, which indicate potential design and implementation deficiencies. This term was first coined by Kent Beck and further explained by Fowler in 1999 [1]. They set out 22 typical code smells and described how to recognize them, such as long method, large class, feature envy. etc. Since then, code smell has become a popular topic in software engineering, attracting continuous attention from both academia and industry.

Over the past decades, numerous approaches for code smell detection and refactoring have been proposed, which can be broadly categorized into metric-based, rule-based, graph-based, and machine-learning-based approaches [2] [3]. The metric-based approach represents the most traditional techniques, and the machine learning approaches have been the most popular research topic in recent studies. Especially, the deep learning approach has received extensive attention from the researchers due to their strong performance [5]-[7].

When applying machine learning to code smell refactoring, the lack of large-scale, high-quality datasets remains a major challenge, especially for deep learning-based approaches. Manual annotation is the most common dataset construction approach; for example, the MLCQ dataset [4] contains over 10,000 developer-reviewed samples. However, manual annotation is costly, time-consuming, and prone to annotator inconsistency, making it difficult to scale. On the other hand, the automatic approaches, such as the *smell-introducing refactoring* proposed by Liu et al. [5], can generate training data from open-source projects, but they still face reliability issues because the quality of automatic generated dataset cannot be guaranteed.

Motivated by the above limitations, we explore the feasibility of combining automated generation techniques with manual process to develop a semi-automatic dataset generation framework. Specifically, we first intentionally create the smelly software entities from the code corpus using pre-defined generation rules. Subsequently, multiple software metrics and rules are applied to categorize both the automatically generated samples and the original code samples into an automatically accepted group and a manually reviewed group. Finally, manual validation is conducted for manually reviewed group based on the established structured review guidelines and an annotation tool developed by us. Compared to the existing code smell dataset generation approaches, our approach leverages automated techniques to generate a large volume of data samples, while incorporating manual verification for ambiguous data samples to ensure dataset quality.

This study is a further extension of the dataset generation approach proposed in the previous studies [6] [7]. Compared to the previous techniques, this study provides a more complete structural process to generate code smell dataset. The details are listed as follows. First, we further improved the semi-automated dataset generation approach to support more code smell. The enhanced approach could generate large amount data samples in three types of code smell data samples: Large Class, Long Method, and Feature Envy. Second, we upgrade all tools in semi-automatic dataset generation approach including generation assistance algorithm and annotation tool. The new annotation tool could be run as a Jetbrain IDE plugin to improve data labeling efficiency. Finally, we utilized this semi-automatic approach to create a large-scale open-source code smell dataset: SACS, covering three widely studied code smells: Long Method, Large Class, and Feature Envy, and each code smell category includes over 10,000 labeled samples.

## II. RELATED DATASET

In this section, we will present several code smell datasets mainly focusing on the dataset generation approach proposed in previous code smell-related studies.

The most commonly used dataset construction approach is the manual approach. A representative example is the MLCQ dataset proposed by Madeyski [4]. It contains approximately 15,000 manually reviewed Java code samples collected from contemporary open-source projects. Each data sample is annotated with four common code smell types—Blob (God Class), Data Class, Feature Envy, and Long Method—and assigned a severity level ranging from none to critical. The labeling process involved 20 professional software developers, and each sample underwent cross-review to ensure annotation reliability. The primary advantage of the manual approach is its potential to guarantee high-quality and accurate labels. However, the cost of manually constructing a code smell dataset is extremely high for several reasons. First, identifying smelly software entities requires substantial development experience, and even expert developers must spend significant time on each data sample. Second, annotation consistency is difficult to maintain, as different developers may give different review results towards ambiguous data samples. Due to these limitations, manually generated datasets are not suitable for machine learning applications that require large amounts of training data.

To address the above limitations, several studies have tried to integrate automatic generation techniques into the dataset generation approach. In Fontana's study [8], to support the machine learning in code smell detection, they included 74 open-source projects from different domains as the data sources. Based on the huge amount of data sources, they used existing refactoring tools as the Advisor to get a candidate list of data samples and manually validated 1,986 data samples for the machine learning task. Although this approach utilized the Advisor to reduce the developer's manually labeling effort, the scale of dataset is still small. And it is hard to get enough positive data samples in dataset. As a result, this approach still does not scale well for large-scale machine learning, especially for deep learning models.

To further improve the automation of dataset generation, the fully automatic dataset generation approach has also been proposed. In Liu's approach [5], they proposed a fully automatic dataset generation approach called *smell-introducing refactoring*. In their approach, they intentionally create a series of smelly software entities as positive data samples from some high-quality open-source projects. In contrast, the other software entities within these projects are considered as non-smelly and extracted as negative data samples. To ensure external behavioral correctness after modification, the automatic correction functionality provided by existing IDEs is applied. This fully automatic strategy significantly improves the efficiency of dataset construction. However, it lost the quality of data samples. As the author pointed out in the paper, the dataset generated in this approach could not guarantee its data quality. First, for positive samples, the approach cannot always generate genuinely smelly software entities, as automatically generated samples do not always exhibit the characteristics of code smells. Second, for negative samples, although high-quality open-source projects are selected as code sources, it cannot be guaranteed that all extracted entities are truly smell-free.

Based on the above discussion, we summarize the existing approaches as follows. First, traditional manual approaches are capable of producing high-quality and reliable data samples; however, they require substantial human effort and time cost. These limitations make it difficult to scale such datasets for machine learning-based approaches. Second, automatic approaches can efficiently generate large volumes of data samples, but they often sacrifice label accuracy and overall data quality.

In this study, we propose a novel semi-automatic dataset generation approach to construct a large-scale high-quality code smell dataset. Specifically, we first pre-defined generation rules for each code smell to generate sufficient positive data samples. Next, we utilized several metrics and formulas to group the data samples into an automatically accepted group and a manually reviewed group, thus the developers could focus the human effort on the ambiguous data samples. Compared with the existing dataset generation approaches, our approach not only preserves the efficiency advantages of automation approaches but also significantly improves dataset quality.

## III. SEMI-AUTOMATIC DATASET GENERATION

### *A. Overview*

The overall workflow of our approach is illustrated in Figure 1. Prior to the dataset generation, we collect several open-source projects as code corpus. Next, in the first step, smelly software entities are intentionally created from the code corpus by pre-defined generation rules. Subsequently, a set of rules is applied to categorize both the automatically generated samples and the original code samples into two groups: *A_Group* and *M_Group*. Data samples in *A_Group* are directly included in the final dataset, whereas the data samples in *M_Group* will undergo manual verification by developers before integration. The manual checking process is facilitated by an annotation tool developed in this study. The details of each step are presented in the following subsections.

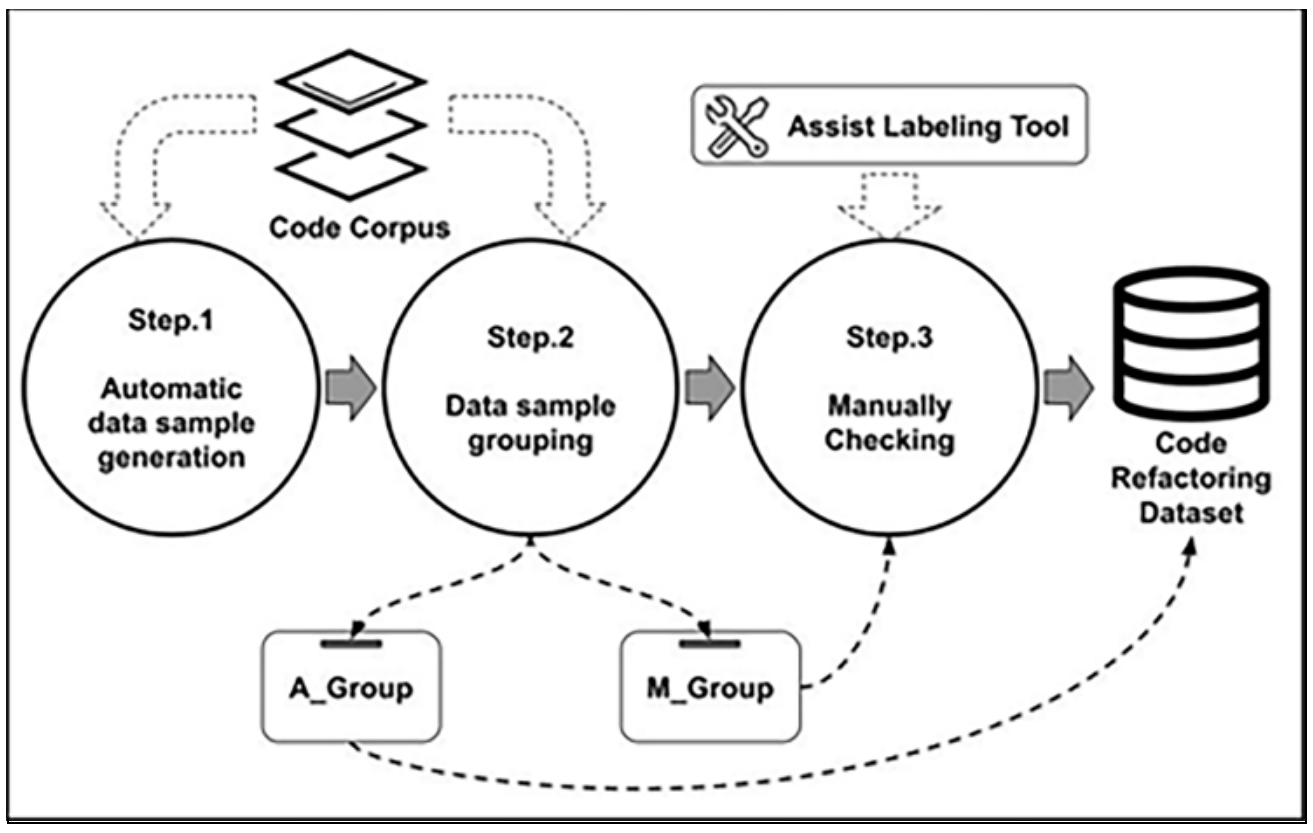

**Figure 1** The overview of dataset generation

### *B. Data Sample Automatic Generation-Step.1*

When constructing a code smell dataset, smelly code samples are often scarce in most of open-source projects. Traditional manual approaches are therefore insufficient to address this challenge. Thus, it is necessary to generate positive data samples automatically. In our approach, smelly software entities are generated from the code corpus according to several predefined rules. The generation rules for each type of code smell are explained as follows.

For long method, we identify three patterns of method invocation as the generation rules to apply method merging accordingly. As shown in Figure 2, the first pattern occurs when a method is directly invoked as a statement. The second pattern arises when a variable within a statement is assigned the return value of another method. The third pattern involves cases where a method is invoked within an expression inside a statement. Once these method-merge opportunities are identified, we perform the merging operation and subsequently resolve any errors introduced during the process.

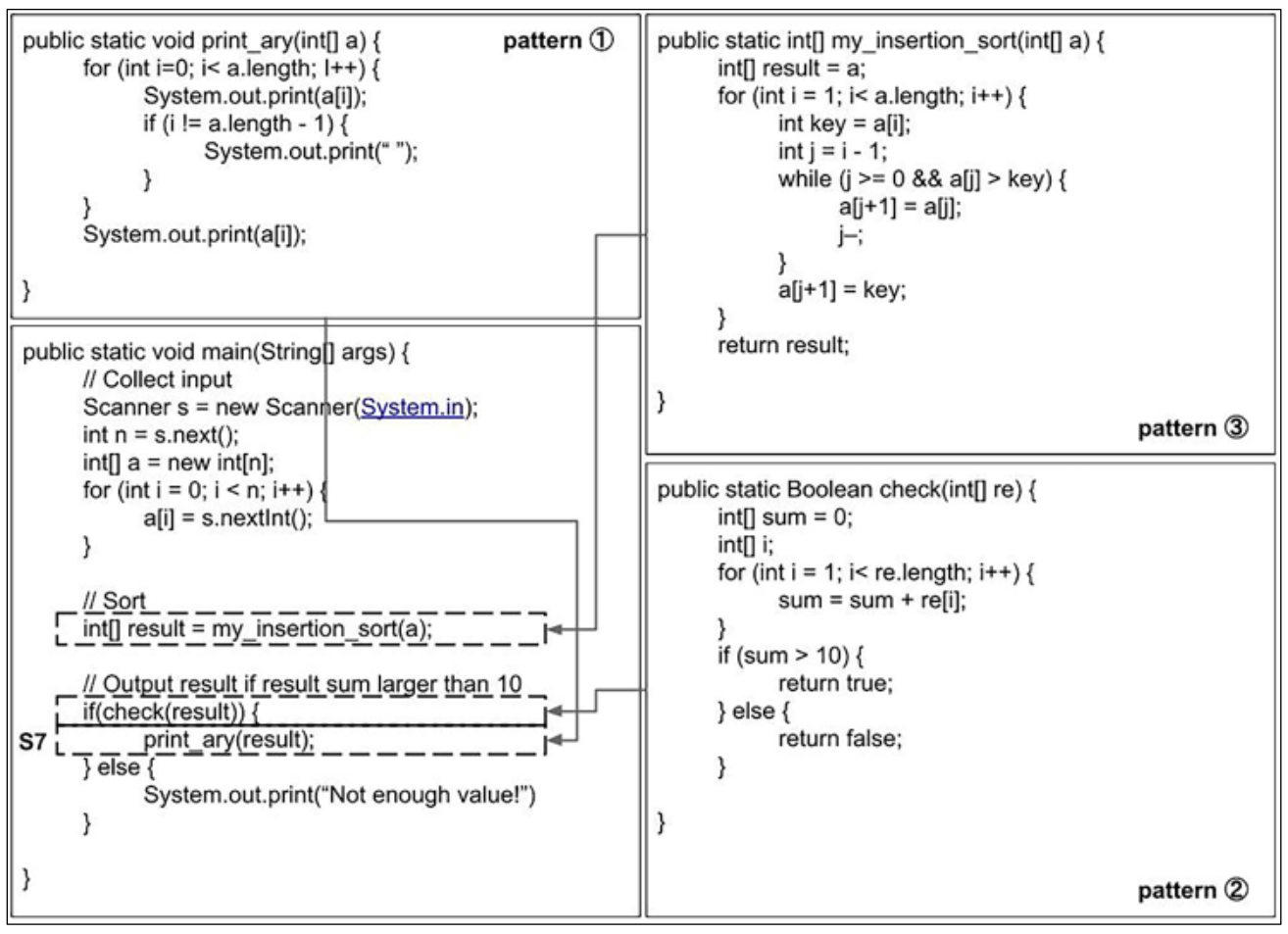

**Figure 2** Long method data sample generation

For example, the code in Figure 2 represents a program that sorts input numbers and prints them when their sum exceeds 10. In Pattern 1, the method *print_ary* is invoked at line S7 of the *main* method. In this case, the two methods exhibit a caller-callee relationship and can be merged by copying all statements from the callee method *print_ary* into the caller method *main*. The parameter *a* in *print_ary* is replaced with the corresponding variable *result* after the merge operation. Methods under Patterns 2 and 3 could be merged in a similar way to generate positive data samples.

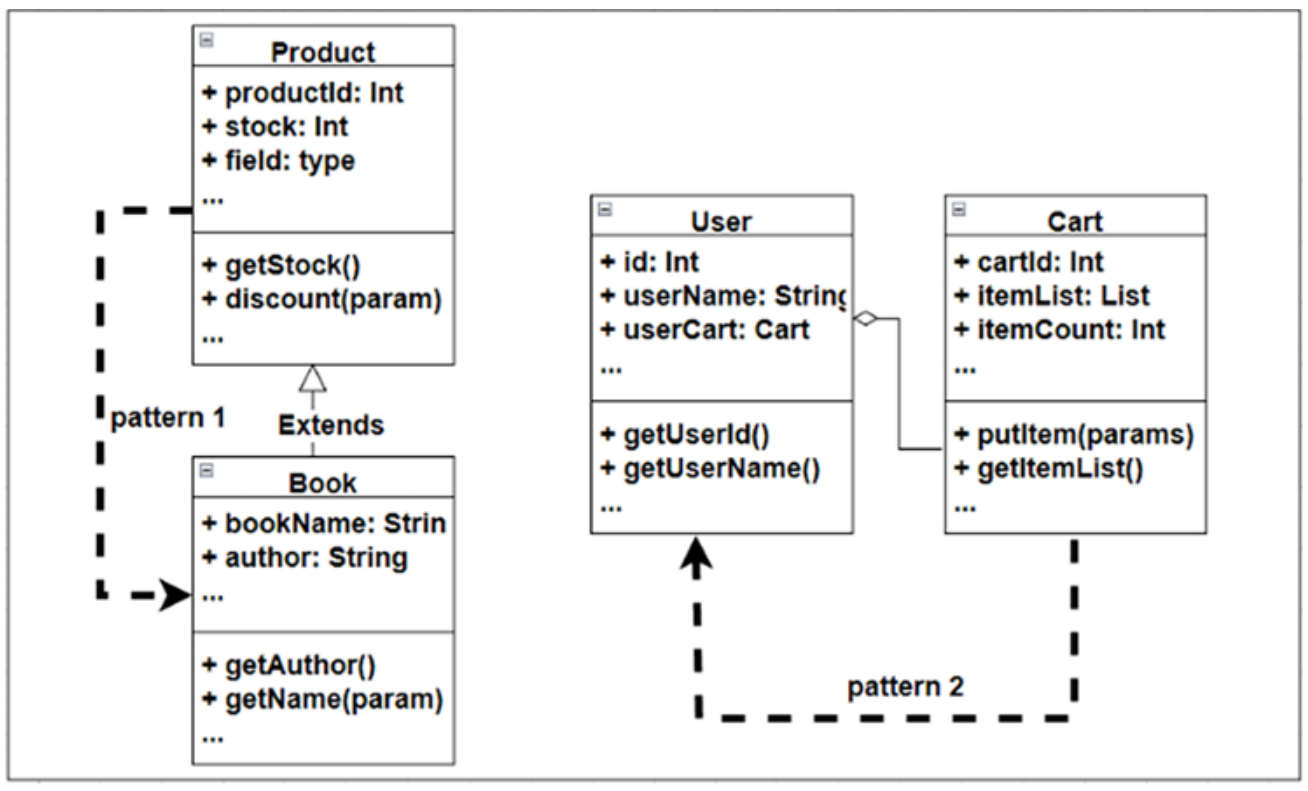

**Figure 3** Large class data sample generation

For large class, we designed two patterns of merging opportunities as shown in Figure 3. These two merging opportunities were designed based on the three refactoring strategies for large classes introduced by Fowler [1]: Extract Class, Extract Superclass, and Replace Type Code with Subclasses. Specifically, the first pattern involves classes with an inheritance relationship, where the parent class can be merged into the child class. For example, in Pattern 1 of Figure 3, the parent class *Product* can be merged into the child class *Book* by copying all methods and fields into the child class. The second pattern involves classes with a usage relationship, where one class is used as a field in another. As illustrated in Figure 3, the class *Cart* is used as a field within the class *User*. In this case, the two classes can be merged by copying all fields and methods from *Cart* into *User*.

For feature envy, we generate data samples by identifying related classes for each method according to three pattern rules and attempting to move the method to the related class. The rules for identifying related classes are illustrated in Figure 4.

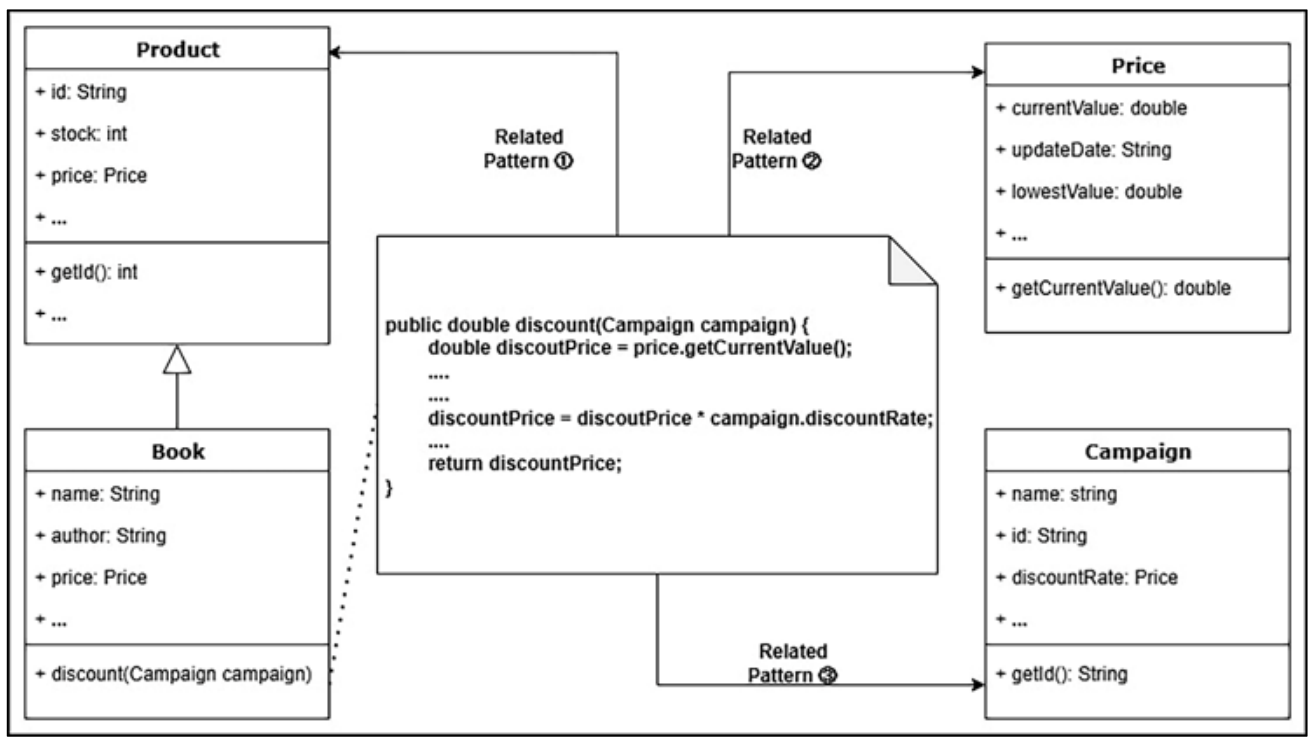

**Figure 4** Feature envy data sample generation

In the first pattern, we determine whether the class of the target method has a parent class. If so, we examine whether any unique fields of the current class are accessed by the target method. If no such fields are used, the method is considered a candidate for relocation to the parent class. In the second pattern, we identify related classes based on property usage. For example, in Figure 4, the method *discount* accesses the property *price* in the *Book* class. Since *price* is defined in the *Price* class, the method *discount* can be relocated from *Book* to *Price*. In the third pattern, related classes are inferred from the parameter list of the target method. For instance, in Figure 4, *campaign* is a parameter of the target method, thus the class *Campaign* is regarded as a related class. In this case, the method discount can be moved to the *Campaign* class.

It should be noted that related classes are restricted to those defined within the project itself; generic language classes or classes defined in external APIs are excluded. If the method relocation succeeds, the target method, along with its original class and target class, is recorded as a feature envy data sample. The reconstruction of methods and target classes differs across patterns. For example, in Pattern 2, the original class *Book* must be added as a field to the target class *Price*, and the invocation relationships in the relocated method must be updated accordingly. In Pattern 3, the parameters of the target method are replaced with references to the source class, and the invocation relationships are likewise modified.

### C. *Data Sample Grouping-Step.2*

To efficiently obtain sufficient high-quality data samples, it is necessary to reduce the workload of human labeling as much as possible. Thus, we must determine which software entities have a higher possibility (or lower possibility) of being smelly, so that we could focus human effort on the ambiguous data samples. To achieve this, we designed several software metrics and rules for each code smell to categorize the data samples into two groups: *A_Group* and *M_Group*. The data samples in the *A_Group* will be directly applied to the final dataset, whereas the data samples in the *M_Group* are included after manual verification.

Table 1. Grouping rules for long method dataset

| Group | Description |
|---|---|
| **A_Group** | 1.The original method with the LOC in PR1(<15)<br>2.The merged method with the LOC in PR3(>30) |
| **M_Group** | 1.The original method which LOC in PR2 (15 to 30) or PR3 (>30)<br>2.The merged method with the LOC in PR1(<15) or PR2 (15 to 30) |

For long method, we used the LOC as the baseline metric and set up three possibility-ranges. Each possibility-range represents the different possibility that the method inside it could be considered as long method. The first range (PR1), covering methods whose LOC falls between 0 and the threshold *MinTv*, indicates a low likelihood of being smelly. The second range (PR2), spanning from *MinTv* to *MaxTv*, represents methods that have a moderate possibility of being long methods. The final range (PR3), which includes methods whose LOC exceed *MaxTv*, corresponds to a high likelihood of being long method. The threshold values were set to *MaxTv* = 30 and *MinTv* = 15 as discussed in the previous study [6]. The possibility-ranges are then used to group the data samples according to the rules specified in Table 1.

For large class, the possibility-ranges are created by three baseline metrics: LOC, Number of Methods (NOM), and Number of Attributes (NOA). The PR1 of large class represents classes with LOC, NOM, and NOA all fall below the minimum thresholds (*MinTv*) , suggesting a low likelihood of being large classes. PR3 includes classes whose LOC, NOM, and NOA all exceeding their respective maximum thresholds (*MaxTv*), indicating a high likelihood of being large classes. PR2 covers all remaining cases, representing classes with a moderate possibility of being large class.

Table 2. Grouping rules for large class dataset

| Group | Rules |
|---|---|
| **A_Group** | 1.The original class in PR1<br>2.The merged class in PR3 |
| **M_Group** | 1.The original class in PR2 or PR3<br>2.The merged class in PR1 or PR2 |

The values of *MaxTv* and *MinTv* for the large class depend on both the target programming language and the characteristics of the projects in the code corpus. Therefore, it is necessary to examine the specifications of the target language or collect statistical information on the relevant metrics within the corpus before determining appropriate threshold values. In this study, based on an empirical survey of the code corpus, the *MaxTv* are set to 10 for NOA, 10 for NOM, and 130 for LOC, while the *MinTv* are set to 5 for NOA, 7 for NOM, and 70 for LOC. Subsequently, the data samples were grouped according to the rules in Table 2. Further detail could also be found in the previous study [7].

For feature envy, we employed a custom metric, Number of Foreign Data Invocation (NFDI) to set up possibility-ranges, inspired by the ATFD proposed by Lanza [9]. Specifically, NFDI measures the frequency with which attributes or methods from external classes are invoked within a given target method. For the PR1, it includes methods whose NFDI values fall between 0 and *MinTv*, indicating a low likelihood of being feature envy. For the PR2, it covering the methods whose NFDI values from *MinTv* to *MaxTv*, represents methods with a moderate possibility of being feature envy. For the PR3, it includes methods whose NFDI values exceed *MaxTv*, indicating a high likelihood of feature envy.

Table 3. Grouping rules for feature envy dataset

| Group | Rules |
|---|---|
| **A_Group** | 1.The original method in PR1<br>2.The moved method in PR3 |
| **M_Group** | 1.The original method in PR2 and PR3<br>2.The moved method in PR1 and PR2 |

The values of *MaxTv* and *MinTv* for feature envy were also determined based on a survey of code corpus data samples. We observed that the methods affected by feature envy typically exhibit an NFDI between 2 and 5, whereas normal methods generally fall within the range of 0 to 5. Consequently, we set *MaxTv* to 5 and *MinTv* to 2, to divide the data samples, according to the rules in Table 3.

### *D. Manually Checking-Step.3*

As described above, the final step of our approach is to manually check the data samples in *M_Group*. For each code smell, the manual verification process encompasses two important tasks. First, the developer determines whether a given data sample is a smelly software entity or not. Second, if the software entity is identified as smelly, the developer specifies the appropriate refactoring action. For instance, which lines should be extracted from a long method, which methods should be extracted from a large class, or to which related target class a feature envy method should be moved. To support accurate annotation, we supply explicit refactoring guidelines for each code smell, thereby promoting both efficiency and consistency in the annotation process. Detailed guidelines for each code smell are provided as follows.

For long method, developers will mark the data samples according to the following guidelines: Is the target method hard to read? Is the target method accessing too many attributes or other methods that may reduce the maintainability? Does the target method have multiple functions or too many parameters, which may reduce the reusability? If the target method is a long method, which lines should be extracted from this method?

For large class, developers will mark the data sample according to the following guidelines: Does the class have too many lines of code? Does the class have too many fields? Does the class have too many complex methods? Does the class have class extraction opportunities that may reduce the reusability of the target class? Does the class have too many responsibilities, which may reduce the maintainability of the target class? If the target class is a large class, which method should be extracted from the target class?

For feature envy, developers will mark the data sample according to the following guidelines: Does the method frequently call from another? Does the method frequently access another class? Does the method rarely use attributes in its own class? Does the method seem more cohesive with another class semantically? If the target method is identified as feature envy, which class should it be moved to?

The annotation process was supported by a custom tool developed as a plugin for JetBrains IDEs. It enables efficient in-IDE annotation, facilitating large-scale code smell labeling through remote collaboration, and is publicly available at: https://github.com/Bankzhy/sce_exp.git.

## IV. OPEN-SOURCE DATASET

Following the above dataset generation approach, we created a large-scale open-source code smell dataset. We collected 16 open-source java projects as the code corpus including: JEdit [11], RxJava [12], Junit4 [13], Mybatis3 [14], Netty [15], Gephi [16], Plantuml [17], Groot [18], MusicBot [19], Traccar [20], Jgrapht [21], Libgdx [22], Freeplane [23], Jsprit [24], Open Hosipital [25], and OpenRefine [26]. We applied semi-automatic generation techniques to the first ten projects as training datasets, while using the remaining six projects as evaluation datasets, mainly based on manual data annotation. The whole dataset including over 13,000 positive data samples and over 70,000 negative data samples. The overview of this dataset is shown in Table 4, and it has been released in https://github.com/Bankzhy/GCSM_Dataset.git.

Table 4. Dataset Overview

| | Long Method | Large Class | Feature Envy |
|---|---|---|---|
| **Positive** | 5,120 | 3,376 | 4,726 |
| **Negative** | 49,991 | 9,722 | 12,578 |
| **Total** | 55,111 | 13,098 | 17,304 |

The dataset described above includes both training and evaluation sets. The training dataset was generated using the semi-automated approach described in Section III. During the automatic sample generation process, an auxiliary program developed by us was employed to support the workflow. The program first parses all target projects into abstract syntax tree. It then generates candidate samples automatically and divides them into two groups according to the rules described in Section III-C. In the subsequent manual validation phase, three experienced Java developers, each with more than eight years of programming experience, were recruited to perform the labeling tasks. The validation process was facilitated by the annotation tool described in Section III-D. Ultimately, we constructed a training dataset comprising more than 20,000 samples. The distribution is presented in Table 5.

Table 5 Training dataset overview

| | Long Method | Large Class | Feature Envy |
|---|---|---|---|
| **Positive** | 4,298 | 3,119 | 4,253 |
| **Negative** | 4,298 | 3,119 | 4,253 |
| **Total** | 8,596 | 6,238 | 8,506 |

Table 6 Evaluation dataset overview

| | Long Method | Large Class | Feature Envy |
|---|---|---|---|
| **Positive** | 822 | 257 | 473 |
| **Negative** | 13,596 | 3,197 | 3,783 |
| **Total** | 14,418 | 3,454 | 4,256 |

For the evaluation dataset, we only applied manually checking phases for all classes and methods. As results, the evaluation dataset consisted of over 20,000 data samples, and the distribution of the evaluation dataset is shown in Table 6. This open-source dataset has been applied to practical code smell–related tasks, and its performance across existing approaches is reported in our study [10].

## V. CONCLUSION

This study proposed a semi-automatic approach to generate a large-scale, high-quality code smell dataset. It first applied rule-based patterns to produce candidate positive samples, and then divided all data samples into automatically accepted and manually reviewed groups using predefined metrics. To ensure labeling quality, detailed guidelines and a custom annotation tool are created to support manual validation. Using this approach, we built the open-source SACS dataset, covering Long Method, Large Class, and Feature Envy, each with over 10,000 labeled samples, providing valuable support for future research on code smell detection and refactoring.

Despite the promising results by our approach, we acknowledge that a small number of erroneous samples may exist among the dataset due to the limitations of the grouping rules. Nevertheless, careful threshold setting and the established possibility-range support the overall reliability, which makes the proposed data generation approach still be considered trustworthy. Moreover, part of the dataset was manually annotated by three developers, and despite detailed guidelines, some subjective inconsistencies were unavoidable. While cross-validation could reduce such bias, it was not adopted due to time cost in this stage. In future work, we will further scale up the dataset by engaging more developers and to generalize the proposed approach to more code smell types.